\documentclass[letterpaper, 10 pt, conference]{ieeeconf}

\usepackage{graphicx}
\usepackage{amsmath}
\usepackage{balance}
\usepackage{multirow}
\usepackage{cite}
\usepackage{stmaryrd}
\usepackage[hidelinks]{hyperref}

\IEEEoverridecommandlockouts         
\newcommand\smallimagewidth{2.75}
{\renewcommand{\arraystretch}{0.9}   
\newcommand{\floor}[1]{\lfloor #1 \rfloor}

\title{\LARGE \bf
	Robust Audio-Based Vehicle Counting in\\Low-to-Moderate Traffic Flow
}

\author{Slobodan Djukanovi\'{c}$^{1}$, Ji\v{r}\'{i} Matas$^{1}$ and Tuomas Virtanen$^{2}$%
	\thanks{Slobodan Djukanovi\'{c} was supported by the OP RDE programme under the project International Mobility of Researchers MSCA-IF III at CTU in Prague No. CZ.02.2.69/0.0/0.0/19\_074/0016255.}%
	\thanks{$^{1}$Slobodan Djukanovi\'{c} and Ji\v{r}\'{i} Matas are with Department of Cybernetics, Faculty of Electrical Engineering, Czech Technical University in Prague, Czech Republic {\tt\small \{djukaslo,matas\}@fel.cvut.cz}.}%
	\thanks{$^{2}$Tuomas Virtanen is with the Audio Research Group, Tampere University, Finland {\tt\small tuomas.virtanen@tuni.fi}.}%
}

\begin{document}
		
	\maketitle
	\thispagestyle{empty}
	\pagestyle{empty}
	
	\begin{abstract}
	The paper presents a method for audio-based vehicle counting (VC) in low-to-moderate traffic using one-channel sound. We formulate VC as a regression problem, i.e., we predict the distance between a vehicle and the microphone. Minima of the proposed distance function correspond to vehicles passing by the microphone. VC is carried out via local minima detection in the predicted distance. We propose to set the minima detection threshold at a point where the probabilities of false positives and false negatives coincide so they statistically cancel each other in total vehicle number. The method is trained and tested on a traffic-monitoring dataset comprising $422$ short, $20$-second one-channel sound files with a total of $ 1421 $ vehicles passing by the microphone. Relative VC error in a traffic location not used in the training is below $ 2 \%$ within a wide range of detection threshold values. Experimental results show that the regression accuracy in noisy environments is improved by introducing a novel high-frequency power feature.
	\end{abstract}
	
	\section{Introduction}
	Traffic monitoring (TM) is used to collect data about the use and performance of roadway systems. The TM data include estimates of vehicle count, flow rate, vehicle speed, vehicle length and weight, vehicle class and identity via the registration plate \cite{won2019intelligent}. Traffic analysis carried out with the collected TM data enables better use of the roadway systems (e.g., enables drivers to be better informed about traffic congestion and parking possibilities), law enforcement (speeding vehicles, dangerous driving, detection of stolen vehicles), prediction of future transportation needs, and the overall improvement of transportation safety.
	
	Based on the sensor type they use, current TM technologies are classified as intrusive, non-intrusive, or off-roadway \cite{won2019intelligent,balid2017intelligent}. Intrusive sensors are embedded in the road and they include magnetic detectors, vibration sensors, piezoelectric sensors and induction loops. Non-intrusive technologies imply mounting sensors overhead on roadways or roadsides and they include cameras, infrared, ultrasonic, magnetic and acoustic sensors, Laser Infrared Detection and Ranging (LIDAR), and Wi-Fi transceivers. The off-roadway systems use mobile sensors installed on aircrafts or satellites.
	
	Vision-based TM systems have several advantages with respect to the other ones. First, they provide rich vehicle-based information such as visual features, vehicle geometry and precise vehicle path. Second, a single sensor is sufficient for detecting and classifying vehicles in multiple lanes. Third, installing cameras to monitor roadways is cheaper and less disruptive than installing sensors in intrusive systems. However, there are many challenges to implement vision-based TM systems, such as partial occlusion, shadows and illumination variation. In addition, video processing for the TM purpose is computationally expensive and time-consuming \cite{won2019intelligent,morris2008survey}.
	
	Audio-based TM offers numerous important advantages with respect to the vision-based one \cite{won2019intelligent}. Microphones are less expensive, consume less energy and require less storage space than cameras. They are not affected by visual occlusions and lighting conditions. They are easier to install and maintain, and have low wear and tear. Finally, microphones are less distractive for drivers\footnote{Drivers change their behaviour when they see a camera since they mistake it for a radar.}.
	
	This paper addresses vehicle counting (VC) in low-to-moderate traffic flow using one-channel sound. To this aim, we propose a vehicle-to-microphone distance function whose minima correspond to vehicles passing by the microphone. The distance function is predicted using regression and VC is carried out via local minima detection in the predicted distance. We propose to set detection threshold so that the false positives and the false negatives statistically cancel each other in total vehicle count. The obtained counting error in a traffic location not included in training is below $ 2 \%$ within a wide range of detection threshold values. In addition to standard acoustic features, we use a novel feature obtained by integrating a high-frequency spectrum of the considered sound signal. Experimental results show that the regression accuracy in noisy environments is improved by the proposed feature. The method is trained and tested on a TM dataset collected for the purpose of this research.
	
	\section{Related work}\label{RelatedWork}
	One of the first attempts to use the sound for automatic estimation of traffic density is presented in \cite{kato2005attempt}. The approach recognizes temporal variations that appear in the power signal when vehicles pass by an observation point. Vehicles are detected based on the state transitions of a hidden Markov model which models global temporal variations of the power signal. Overlapping of the vehicles' sounds is resolved by separately processing channels from a stereo microphone.
	
	In \cite{george2013vehicle} and \cite{george2013exploring}, vehicle detection based on the short-time energy is performed. From the smoothed logarithmic energy, VC is carried out by detecting peaks via a peak picking algorithm. Spurious peaks due to horns are eliminated by spectral filtering of the sound signal. Features extracted from this signal are used to classify the vehicles into heavy, medium and light categories.
	
	In \cite{li2017auto++}, an unsupervised approach for detecting approaching cars, referred to as Auto++, was proposed. The authors recognized the challenges of the lack of temporal structure of vehicular noise and complex acoustic noise characteristics, which can be further emphasized by low-quality sound recording devices. To address these challenges, a novel robust feature is proposed, top-right frequency, which represents maximum frequency whose power reaches a certain threshold. The added value of this approach is possibility to be implemented on devices with low CPU and memory requirements, such as smartphones, allowing its users to detect oncoming vehicles.
	
	The use of microphone arrays for VC and motion direction estimation is considered in \cite{severdaks2013vehicle} and \cite{zu2017vehicle}. The authors in \cite{severdaks2013vehicle} use sound delay in three microphones for these tasks. Small-aperture microphone array is used in \cite{zu2017vehicle} and spatial coherence is used to select the useful bands for determining the vehicle direction, counting vehicles and estimating their moving direction.
	
	\section{Proposed vehicle counting}\label{ProposedMethod}
	
	\subsection{Assumptions}
	
	The road situation considered in our research is limited to one-direction two-lane roads. No assumptions are made regarding the speed of the vehicles, nor it is estimated. The recorded sound is defined as a mixture of the sound produced by vehicles to be counted and the environmental noise, which can include the sound of other vehicles not to be counted (due to vicinity of other roads), weather conditions (no wind to moderate wind, no rain) as well as sound reflections of vehicles to be counted (e.g., due to walls by the road).
	
	\subsection{Dataset collection, preprocessing and splitting}
	
\begin{figure}[thpb]
		\centering
		\begin{tabular}{ccc}
			\includegraphics[width=\smallimagewidth cm, keepaspectratio]{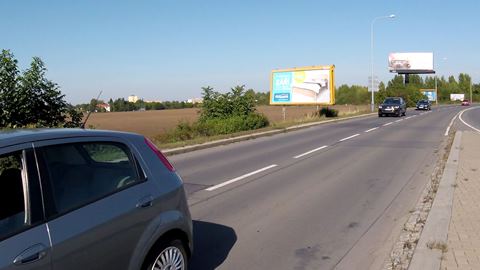}&
			\includegraphics[width=\smallimagewidth cm, keepaspectratio]{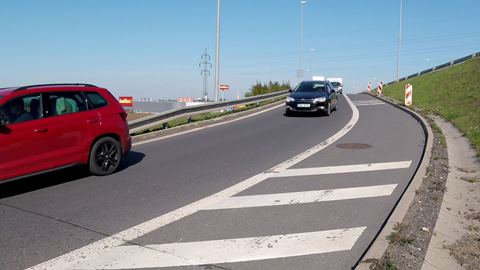}&
			\includegraphics[width=\smallimagewidth cm, keepaspectratio]{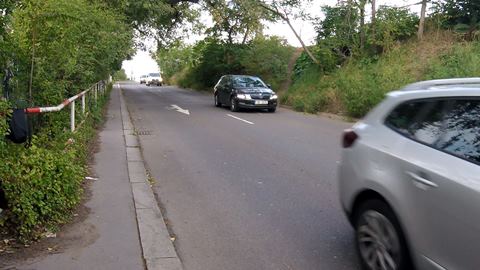}\\
			\includegraphics[width=\smallimagewidth cm, keepaspectratio]{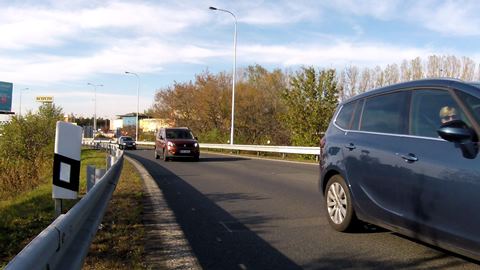}&
			\includegraphics[width=\smallimagewidth cm, keepaspectratio]{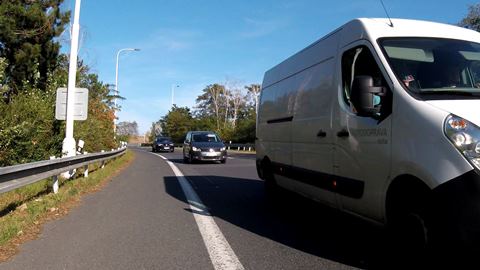}&
			\includegraphics[width=\smallimagewidth cm, keepaspectratio]{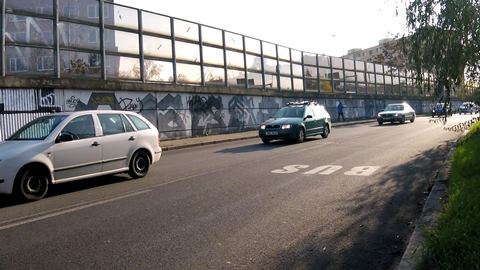}\\
		\end{tabular}
		\caption{Data acquisition at six different locations in Prague.}
		\label{Fig1}
	\end{figure}

	A dataset (audio-video recordings of the road traffic) has been recorded using a GoPro Hero5 Session camera. It was installed on a sidewalk, at a safe distance of at least $ 0.5 $ m from the road and at the height of around $ 1.2 $ m. Figure \ref{Fig1} presents sample images from the camera in six different two-lane one-direction roads in Prague, Czech Republic. The roads are located away from the city centre, but are within the innermost tariff zones P, $ 0 $ and B \cite{pob}. We have picked various camera positions (both sides of the road and different angles of the camera with respect to the road) in order not to be sensitive to the actual camera position. The dataset acquisition took place from September to November 2019.
	
	The recorded video material is divided into $ 422 $ short, $ 20 $-second non-overlapping video clips using the Format Factory application (version 4.9.0). The number of video clips per location is $ 15 $, $ 32 $, $ 26 $, $ 112 $, $ 65 $ and $ 172 $ (Fig. \ref{Fig1}, left to right, top to bottom). The sound files ($ 44100 $ Hz sampling rate, WAV format, $ 32 $-bit float PCM) have been extracted from the video clips using Audacity (version 2.3.2), a free open-source application for recording and editing sound. The total material includes $ 1421 $ vehicles passing by the microphone, making an average of $ 303 $ vehicles/hour/lane. 
	
	Annotation data contain the pass-by-microphone times of all vehicles. We record the relative time from the beginning of the file, measured in seconds with a two-decimal precision. Precise annotations are obtained by visual screening, i.e., by identifying a video frame when a vehicle starts to exit the camera view, which is approximately the moment when it is closest to the microphone. Five vehicle classes are considered in the annotation: motorcycles, cars, vans, buses and trucks.
	
	The dataset is split into two parts. The first part, referred to as VC-PRG-1:5\footnote{PRG in VC-PRG-1:5 stands for Prague.}, corresponds to the first five recording locations in Fig. \ref{Fig1} and it consists of $ 250 $ sound files ($20$-second segments) with $ 841 $ vehicles in total. The second part, referred to as VC-PRG-6, corresponds to the sixth location (last image in Fig. 1), it contains $ 172 $ sound files with $ 580 $ vehicles in total. VC-PRG-1:5 will be used to evaluate the proposed method via the standard $ k $-fold cross validation where all five locations will be included in both the training and the testing phase, whereas VC-PRG-6 will be used for evaluation on data collected at a location not considered in the training. The dataset is available for download at
	\url{http://cmp.felk.cvut.cz/data/audio_vc}.
	
	\subsection{Features}\label{Features}
	
	The proposed VC approach is based exclusively on audio cues. When a vehicle is directly in front of the microphone, the energy of the corresponding sound signal will reach a local maximum \cite{kato2005attempt, george2013exploring, george2013vehicle}, i.e., a peak in the energy gives an indication about the presence of a vehicle. However, peaks in energy are also induced by other vehicles (in vicinity of other roads), by wind or some other source. 
	
	The features will be calculated from a sound signal $x(n)$, where $n=1,2,\dots , N$ represents discrete time variable and $ N $ is the signal length. Description of the features follows, along with the implementation details.
	
	\subsubsection{Short-term energy}\label{STE_sec}

	Short-term energy (STE) of $ x(n) $ is calculated as
	\begin{equation}\label{STE}
	\textrm{STE}(m)=\frac{1}{N_w}\sum\limits_{n=1}^{N_w}{x^{2}(n+(m-1)N_h)}, \quad m=1,\dots,M
	\end{equation}
	where $ N_w $ and $ N_h $ represent the window and hop lengths, respectively. The STE length $ M $ equals the number of hops, i.e., $ M=\floor{\frac{N-N_w}{N_h}} + 1$, where $ \floor{\cdot} $ is the round-down operator. In \cite{george2013vehicle}, the authors use Hamming window and an STE smoothing with a lowpass Bessel filter in order to remove high frequency fluctuations.
	
	\subsubsection{Top-right frequency}
	
	Top-right frequency (TRF) \cite{li2017auto++} represents maximal frequency whose power reaches a predefined threshold \textit{T}, i.e.
	\begin{equation}\label{TRF}
	\textrm{TRF}(m)=\mathrm{max}\left(f\llbracket X(m,f)\geq T\rrbracket\right), \quad f \in \left[0,f_{max}\right],
	\end{equation}
	where $ X(m,f) $ represents a time-frequency power distribution of $ x(n) $, $ f $ is the frequency, $ m $ discrete time variable, and $ \llbracket \cdot\rrbracket $ is the Iverson bracket defined as
	\begin{equation}\notag
	\llbracket S\rrbracket=\begin{cases}
		1,	& \quad \text{statement } S \text{ is true}\\
		0, & \quad \text{otherwise}.
	\end{cases}
	\end{equation}
	We will calculate $ X(m,f) $ as the spectrogram, the squared modulus of the short-time Fourier transform (STFT) \cite{stankovic2014instantaneous}.
	
\begin{figure}[thpb]
		\centering
		\includegraphics{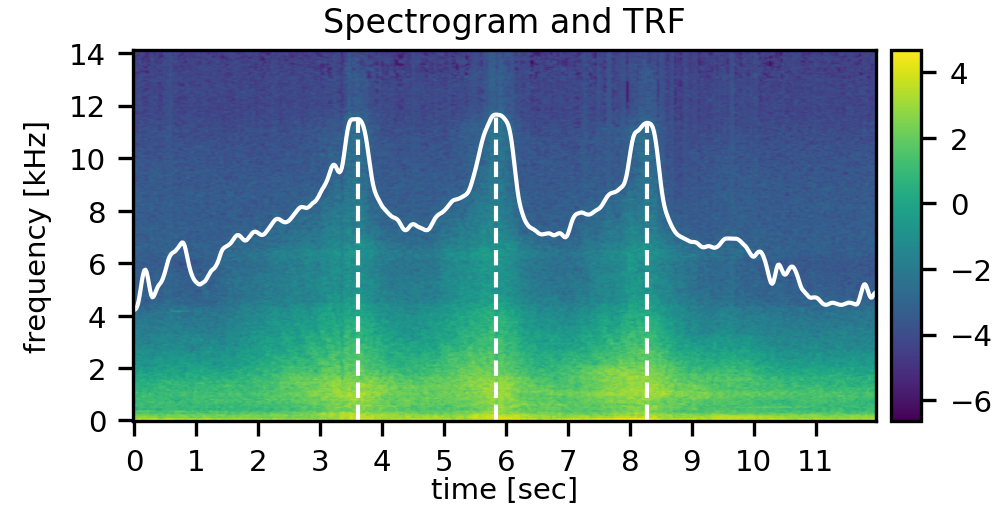}
		\caption{Spectrogram (log-amplitude scale) and time-varying TRF of the signal.}
		\label{Fig2}
	\end{figure}

	As a vehicle gets closer to the microphone, the energy across all frequency components rapidly increases, reaching a maximum when the vehicle passes by the microphone. This trend holds also for the signal's TRF (white solid line in Fig. 2). The TRF maxima (dotted lines in Fig. \ref{Fig2}) correspond to moments when vehicles pass by the microphone.
	
	\subsubsection{High-frequency power}
	
\begin{figure}[thpb]
		\centering
		\includegraphics{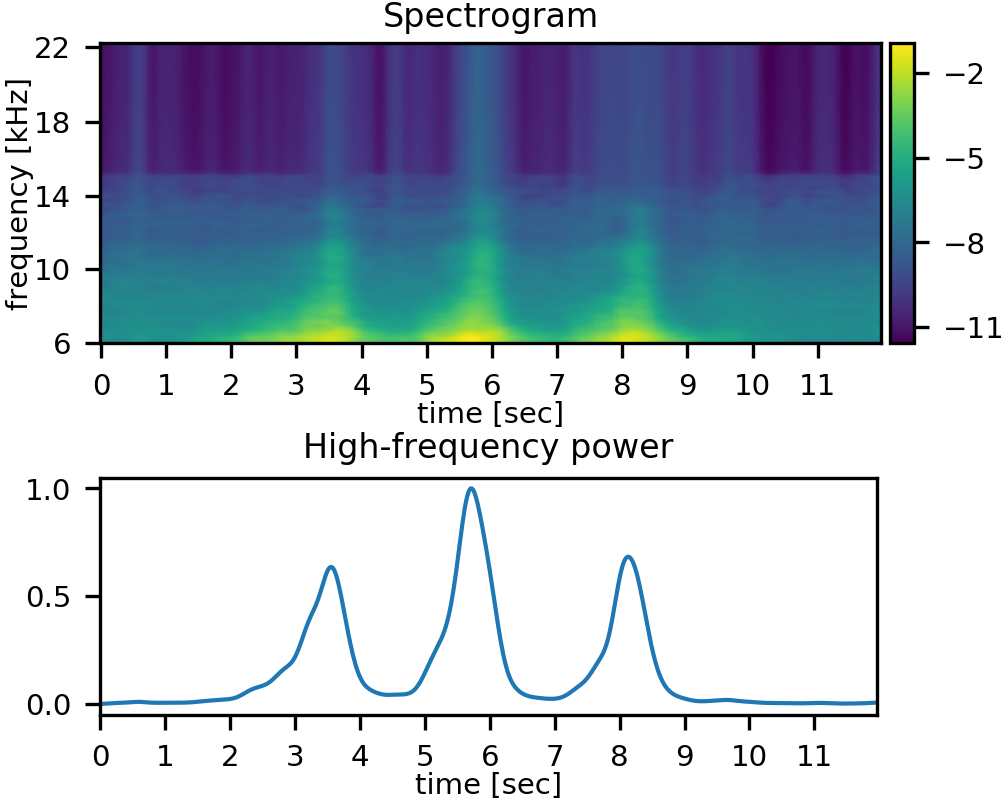}
		\caption{\textit{Top}: High-frequency (above 6 kHz) portion of the spectrogram (log-amplitude scale). \textit{Bottom}: High-frequency power feature.}
		\label{Fig3}
	\end{figure}

	In this paper, we introduce a new VC feature, calculated as the power of a high-frequency portion of the signal spectrum, and hence referred to as \textit{high-frequency power} (HFP):
	\begin{equation}\label{HFP}
	\text{HFP}(m)=\int_{f=f_{min}}^{f_{max}}X(m,f)df,
	\end{equation}
	where $ f_{min} $ and $ f_{max} $ represent minimal and maximal considered frequencies, respectively. $ f_{max} $ can be set to maximum spectral frequency (half of the sampling frequency according to the Nyquist criterion), whereas $ f_{min} $ is set so that a low-frequency part of the spectrum, which includes the background noise, is skipped. Figure \ref{Fig3} (top) presents high-frequency (above $ 6 $ kHz) portion of the spectrogram of the same signal as in Fig. \ref{Fig2}, whereas the bottom plot depicts the proposed HFP feature. The HFP peaks due to vehicles passing by the microphone are much more prominent than the TRF ones.
	
	\subsubsection{Log-mel spectrogram}
	
	Log-mel spectrogram (LMS) represents the short-term power spectrum of a sound signal projected to a reduced set of frequency bands and converted to logarithmic magnitude. It represents the standard acoustic feature for audio analysis tasks \cite{serizel2018acoustic}.
	
	\subsubsection{Implementation of features}
	
	All the features possess the time dimension, i.e., STE, TRF and HFP are 1-D time functions, whereas LMS is a 2-D function with time as one dimension. The TRF, HFP and LMS features are based on STFT, hence their time dimension depends on the number of window positions in the STFT calculation. In this paper, we use $ N_w=4096 $ and $ N_h=0.4N_w=1638 $ samples, which with $20$-second audio files sampled at $ 44100 $ Hz gives the time-length of all features of $ 539 $ samples. In addition, $ N_{mel}=64 $ mel bands are used in the LMS calculation, so this feature's shape is $ 64\times 539 $. Hamming window is used in the STFT calculation. In the HFP feature, $(f_{min},f_{max})=(6000 \textrm{ Hz},22050 \textrm{ Hz})$.
	
	Prior to using the STE, TRF and HFP features, we filter them to eliminate high-frequency oscillations. To that end, we apply two successive moving average (MA) filters, first with the length of $ 11 $, then of $ 5 $ samples. After the filtering, these features are normalized to zero mean and unit variance. Finally, at each time instant $ n $, we do not consider the values of STE, TRF and HFP only at $ n $, but also at $ K $ preceding and $ K $ following values, i.e., at each time instant, we consider $ 2K+1 $ successive values of these features. The boundary values have been handled by quadratic extrapolation\footnote{For example, at $ n=1 $, the first $ K $ out of $ 2K+1 $ values of a feature $ F(n) $ are obtained are extrapolating the values $F(1), F(2), \dots, F(K+1)$.}. In this paper, $ K=10 $. Therefore, the total number of features is $3(2K+1)+N_{mel}=127$.

	\subsection{Proposed method}
	
\begin{figure}[thpb]
		\centering
		\includegraphics{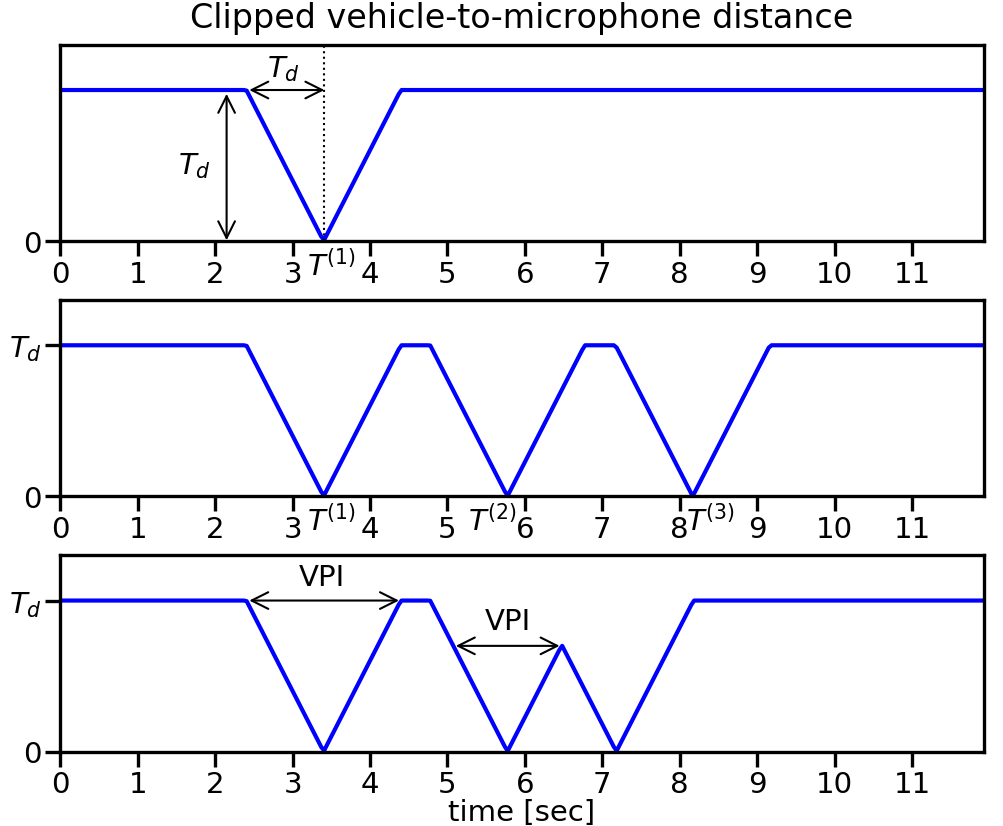}
		\caption{Clipped vehicle-to-microphone distance (\textit{x}- and \textit{y}-axis are in different scale). \textit{Top}: One vehicle. \textit{Middle}: Vehicles are apart by at least $ 2T_d $ seconds. \textit{Bottom}: Vehicles are less than $ 2T_d $ seconds apart.}
		\label{Fig4}
	\end{figure}

	For the VC purpose, we propose to use a function analytically defined as follows:
	\begin{equation}\label{distance}
	d^{(l)}(t)=\begin{cases}
		\left|t-T^{(l)}\right|,	& \quad \left|t-T^{(l)}\right| < T_d\\
		 T_d, & \quad  \text{elsewhere},
	\end{cases}
	\end{equation}
	where $ T^{(l)} $ represents the moment when the \textit{l}-th vehicle passes by the microphone and $ T_d $ is the distance threshold. Function $ d^{(l)}(t) $ has the form of a clipped distance of a vehicle, driven at uniform speed, from the microphone, and is depicted in Fig. \ref{Fig4} (top). Hence, $ d^{(l)}(t) $ will be referred to as \textit{clipped vehicle-to-microphone distance} (CVMD) of the \textit{l}-th vehicle. With multiple vehicles, each vehicle is characterized by a V-shape profile and CVMD takes the general form
	\begin{equation}\label{CVMD}
	D(t)=\min\{d^{(1)}(t),d^{(2)}(t),\dots, d^{(N_{v})}(t)\},
	\end{equation}
	where $ N_{v} $ represents the number of vehicles in the corresponding sound signal. CVMD for the signal considered in Figs. \ref{Fig2} and \ref{Fig3} is presented in Fig. \ref{Fig4} (middle), whereas the case when two vehicles are apart by less than $ 2T_d $ seconds is presented in Fig. \ref{Fig4} (bottom). The value of CVMD is either the distance to the nearest vehicle or $ T_d $.

	\begin{figure}[thpb]
		\centering
		\includegraphics{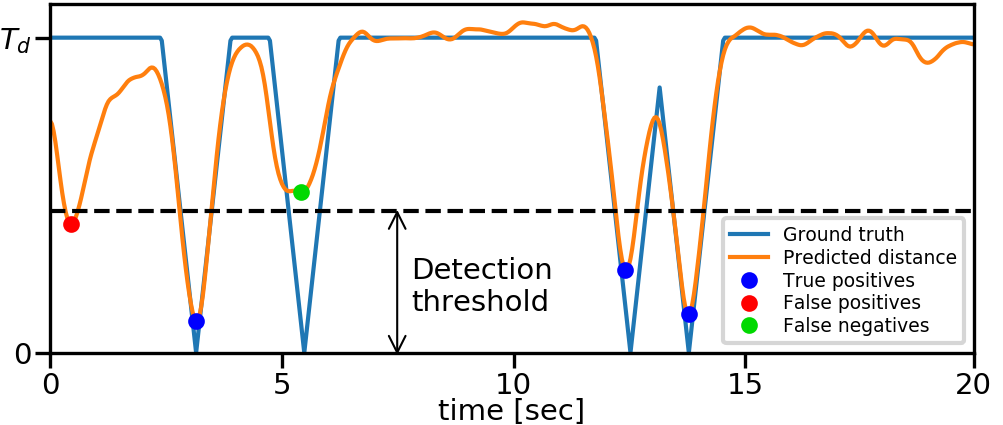}
		\caption{Classification of local minima in the predicted distance.}
		\label{Fig5}
	\end{figure}

	The CVMD function allows us to distinguish between the sounds produced by vehicles passing by the microphone and other sounds for which the V-shape distance profile does not make sense. It is predicted using a regression procedure having as input the features described in Section \ref{Features}. The regression implementation details follow soon. Figure \ref{Fig5} depicts an example of the regression output (orange line) for a given ground truth CVMD (blue line). Our VC approach is to detect local minima in the predicted distance which surpass a detection threshold (dashed line in Fig. \ref{Fig5}). However, not every local minimum surpassing the threshold is associated with a vehicle passing by the microphone (see the red dot in Fig. \ref{Fig5}). An additional requirement for a minimum-to-vehicle association is that a local minimum occurs within the corresponding V-shape profile. To that end, we define \textit{vehicle pass-by interval} (VPI) which will represent a basis for deciding whether a local minimum corresponds to a vehicle passing by the microphone (VPI is presented in Fig. \ref{Fig4} (bottom)). Therefore, we define two criteria which will allow us to classify the detected local minima:
	\begin{description}
		\item[C1:]\label{C1} local minimum surpasses the detection threshold,
		\item[C2:]\label{C2} local minimum occurs within a VPI.
	\end{description}
	
	A local minimum satisfying both C1 and C2 corresponds to a correctly detected vehicle, or a true positive (TP) (blue dots in Fig. \ref{Fig5}). On the other hand, a local minimum satisfying C1 and not C2 corresponds to a false positive (FP) (red dot in Fig. \ref{Fig5}), whereas a local minimum satisfying C2 and not C1 corresponds to a false negative (FN) (green dot in Fig. \ref{Fig5}). The total number of detected vehicles equals the sum of the TPs and the FPs.

	\subsubsection{Implementation details}
	
	In our experiments, for distance threshold we adopt $ T_d=0.75 $ s\footnote{We did not carry out a detailed analysis of the influence of the $ T_d $ value on the method performance. $ T_d=0.75 $ is adopted based on a comparison of performances obtained with a few smaller and a few larger values.}. We apply the support vector machine (SVM) approach for the CVMD regression, more precisely the $ \varepsilon $-support vector regression ($ \varepsilon $-SVR). The free parameters in the model are $ C $ (penalty parameter of the error term) and $ \varepsilon $ (controls the width of the $ \varepsilon $-insensitive zone, used to fit the training data, i.e. determines the accuracy level of the approximated function) \cite{chang2011libsvm}. A grid search on $ C $ and $ \varepsilon $ values yielded the optimal values of $ C_{opt}=1 $ and $ \varepsilon_{opt}=0.05 $. Further reducing the $ \varepsilon $ value would increase the model complexity and lead to overfit \cite{mattera1999support}. 
	
	The SVR is implemented using the \texttt{sklearn.svm.SVR} procedure from the scikit-learn library for machine learning in Python. Peak detection is performed using the \texttt{scipy.signal.find\_peaks} procedure (SciPy v.1.3.0 library for Python), with all parameters set to default values except for the peak prominence set to $ 0.05 $. Prior to peak detection, we filter the SVR results to eliminate high-frequency oscillations (three successive MA filters with lengths $ 7 $, $ 5 $ and $ 3 $).
	
	Method implementation in Python is available for download at	\url{http://cmp.felk.cvut.cz/data/audio_vc}.

	\section{Experiment}\label{Experiment}
	
	The proposed method is evaluated by calculating the TP, FP and FN probabilities, denoted as $ p_{\text{TP}} $, $ p_{\text{FP}} $ and $ p_{\text{FN}} $, respectively. These probabilities are calculated versus detection threshold varying from $ 0 $ to $ T_d $ in steps of $\Delta T_d=1\%T_d $. In addition to these probabilities, we will calculate {\it{normalized area under the curve}} (NAUC) $ p_{\text{TP}} $ as follows:
	\begin{equation}\label{NAUC}
	\text{NAUC}=\frac{\sum_{i=0}^{I_{max}-1}p_{\text{TP}}(i\Delta T_d)\Delta T_d}{T_d},
	\end{equation}
	where $ p_{\text{TP}}(i\Delta T_d) $ represents the value of $ p_{\text{TP}} $ obtained for detection threshold $ i\Delta T_d $, whereas $ I_{max} $ is the number of detection thresholds, in our case $ 100 $. Clearly, the bigger NAUC the better, and $ \text{NAUC}=1 $ is the optimal value.
	
	The proposed method requires selecting the detection threshold. The natural choice would be a point where $ p_{\text{FP}} $ and $ p_{\text{FN}} $ coincide since then, in statistical sense, the FPs and the FNs cancel each other so that the total number of detected vehicles equals the true number of vehicles. Therefore, another evaluation metric will be introduced, {\it{equal false probabilities}} (EFP), as a value $ \text{EFP}=p_{\text{FP}}=p_{\text{FN}}$. Having in mind the discrete nature of detection threshold, EFP will be calculated as 
	\begin{equation}\label{EFP}
		\text{EFP}=p_{\text{FP}}(I_{min}\Delta T_d),
	\end{equation}
	where
	\begin{equation}\label{Imin}
		I_{min}=\underset{i}{\mathrm{argmin}}\ |p_{\text{FP}}(i\Delta T_d)-p_{\text{FN}}(i\Delta T_d)|. 
	\end{equation}
	As opposed to NAUC, the smaller EFP the better. Ideally, $ \text{EFP}=0 $.
	
	Finally, let us define \textit{relative VC error} (RVCE) as
	\begin{equation}\label{RVCE}
	\text{RVCE}=\frac{\left|N_{v}^{true}-N_{v}^{est}\right|}{N_{v}^{true}} \times 100 \,[\%],
	\end{equation}
	where $ N_{v}^{true} $ and $ N_{v}^{est} $ represent the true and the estimated number of vehicles in the considered dataset.

	\subsection{Regression with all features}\label{RegAllFeat}
	
\begin{figure}[thpb]
		\centering
		\includegraphics{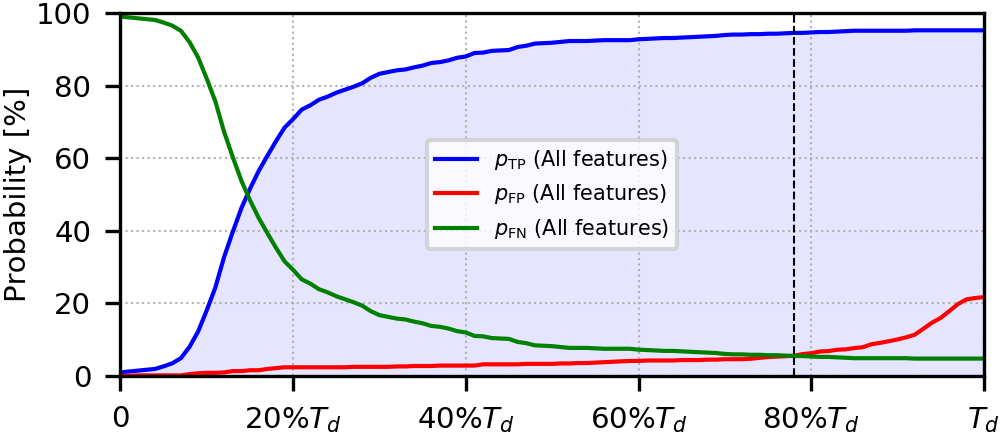}
		\caption{TP, FP and FN probabilities. Dashed line corresponds to detection threshold where $p_{\text{FP}}=p_{\text{FN}}$.}
		\label{Fig6}
	\end{figure}

	Figure \ref{Fig6} depicts the TP, FP and FN probabilities obtained after regression with all features. With low detection thresholds, criterion C1 is rarely met, resulting in low $ p_{\text{TP}} $ and high $ p_{\text{FN}} $. An opposite situation is obtained for high detection thresholds, when majority of detected local minima surpass the threshold. In that case, $ p_{\text{FP}} $ also rises since numerous false minima will also surpass the threshold. The NAUC and EFP values are presented in Table \ref{TableResults1} (Setup I). In addition to EFP, we report the minimal absolute difference between $p_{\text{FP}}$ and $p_{\text{FN}}$, i.e.
	\begin{equation}\label{deltaEFP}
	\Delta_{\text{EFP}}=|p_{\text{FP}}(I_{min}\Delta T_d)-p_{\text{FN}}(I_{min}\Delta T_d)|,
	\end{equation}
	with $ I_{min} $ defined in (\ref{Imin}).
	
	The TP probability is high for a wide range of detection threshold values. For example, $ p_{\text{TP}}$ exceeds $ 90\% $ and $ 94\%$ for thresholds above $ 46\% T_d $ and $ 71\% T_d $, respectively. Detection threshold corresponding to EFP ($ p_{\text{FP}}=5.52 \%$ and $p_{\text{FN}}=5.45 \%$) equals $ 78\% T_d $ and is shown in Fig. \ref{Fig6} (dashed line). For this threshold value, $p_{\text{TP}}=94.55\%$.
	
	\subsection{Regression with feature combinations - Ablation study}\label{RegFeatComb}
	
\begin{figure}[thpb]
		\centering
		\includegraphics[width=0.48\textwidth]{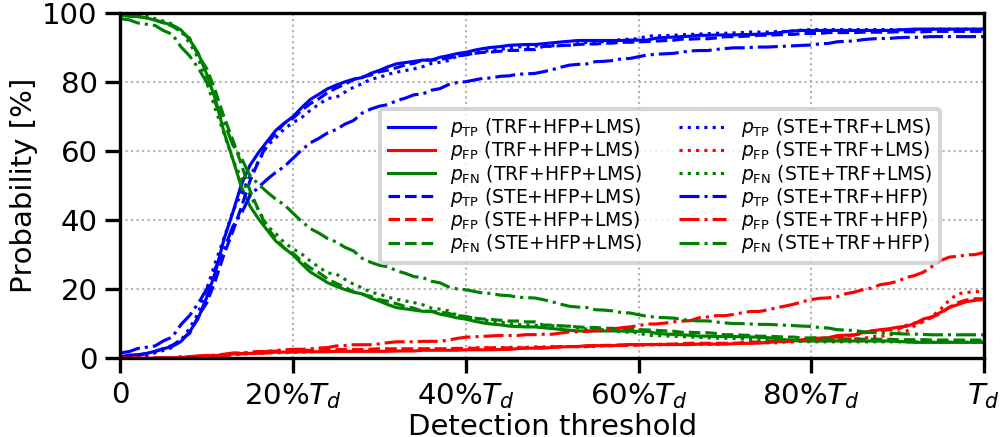}
		\caption{TP, FP and FN probabilities for three-feature combinations.}
		\label{Fig7}
	\end{figure}

	To see how the exclusion of features affects the regression performance, we repeat the experiment from Section \ref{RegAllFeat} with different feature combinations. More precisely, we consider four sets of features, each obtained by leaving out exactly one feature from the feature set. The corresponding probability curves are presented in Fig. \ref{Fig7}, whereas the NAUC and EFP metrics are given in Table \ref{TableResults1} (Setup II). Clearly, the exclusion of LMS (dash-dot line in Fig. \ref{Fig7} and the last combination in Setup II in Table \ref{TableResults1}) affects performance the most. There are no significant differences in performances of the other three combinations. TRF+HFP+LMS is characterized by the biggest NAUC of all feature combinations in Table \ref{TableResults1}.
	
\begin{figure}[thpb]
		\centering
		\includegraphics[width=0.48\textwidth]{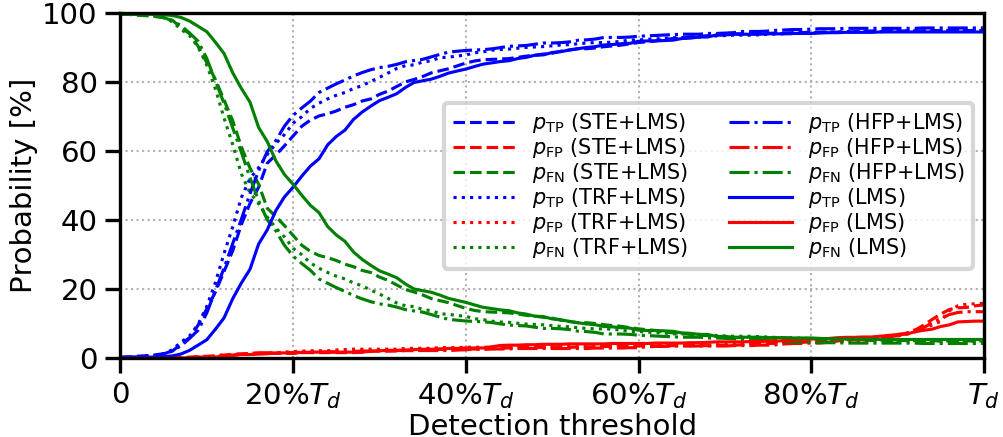}
		\caption{TP, FP and FN probabilities for combinations of LMS with other features and LMS alone.}
		\label{Fig8}
	\end{figure}

	Having identified the LMS as the most critical feature, we repeat the experiment with i) two-feature combinations, one of them being LMS, and ii) LMS alone. The results are presented in Fig. \ref{Fig8} and Table \ref{TableResults1} (Setups III and IV). The best performance in terms of both NAUC and EFP is obtained with HFP+LMS. Moreover, this feature combination outperforms all other combinations in Table \ref{TableResults1} in terms of EFP. Finally, the NAUC performance of LMS alone is notably worse than those of other combinations, which can be also seen in Fig. \ref{Fig8} (solid line).
	

	\begin{table}[h]
		\setlength{\tabcolsep}{0.5em} 
		\renewcommand{\arraystretch}{1.1}
		\centering
		\caption{NAUC and EFP metrics for the VC method trained and tested on VC-PRG-1:5. $ \Delta_{\text{EFP}}=\min|p_{\text{FP}}-p_{\text{FN}}|$.}
		\begin{tabular}{clcc}
			\hline\hline
			\multicolumn{1}{c}{Setup \#} & \multicolumn{1}{c}{Features} & 
			\multicolumn{1}{c}{NAUC} & \multicolumn{1}{c}{EFP [\%] ($ \Delta_{\text{EFP}}$ [\%])} \\
			\hline\hline
			I & All features & $0.773$ & $5.52$ ($0.07$) \\
			\hline
			\multirow{4}{*}{II} & TRF+HFP+LMS & $\mathbf{0.775}$ & $5.09$ ($0.06$) \\
								& STE+HFP+LMS & $0.766$ & $5.68$ ($0.23$) \\
								& STE+TRF+LMS & $0.770$ & $4.99$ ($0.10$) \\
								& STE+TRF+HFP & $0.721$ & $10.91$ ($0.06$) \\
			\hline
			\multirow{3}{*}{III} & STE+LMS & $0.748$ & $5.49$ ($0.09$) \\
								 & TRF+LMS & $0.764$ & $5.40$ ($0.02$) \\
								 & HFP+LMS & $0.772$ & $\mathbf{4.43}$ ($0.09$) \\
			\hline
			IV & LMS & $0.719$ & $5.75$ ($0.07$) \\
			\hline\hline
		\end{tabular}
		\label{TableResults1}
	\end{table}

	\subsection{Method generalization}\label{Generalization}
	
\begin{figure}[thpb]
		\centering
		\includegraphics[width=0.48\textwidth]{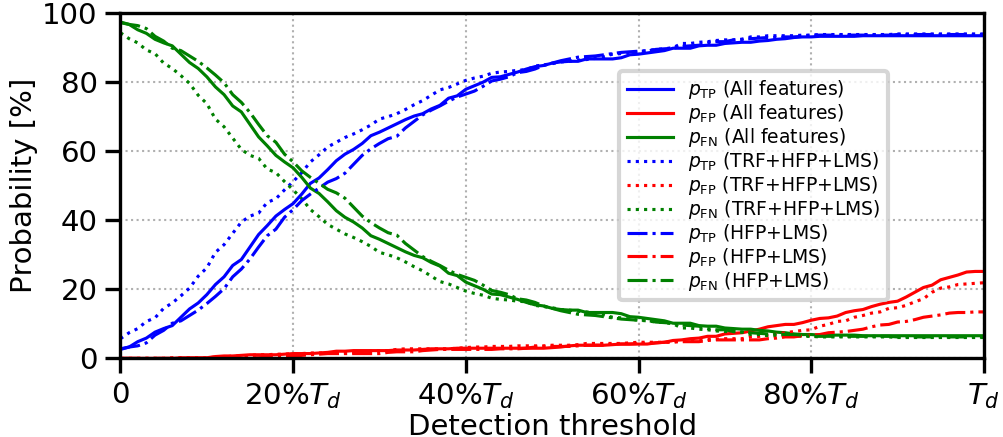}
		\caption{TP, FP and FN probabilities for the case when different datasets are used for training and testing.}
		\label{Fig9}
	\end{figure}

	In this experiment, the method is trained on VC-PRG-1:5 and tested on VC-PRG-6. The location corresponding to VC-PRG-6 differs from those of VC-PRG-1:5 in a road-noise barrier erected on the other side of the road, which causes undesirable acoustic reflection. We consider all features and two best performing combinations (TRF+HFP+LMS and HFP+LMS) from Table \ref{TableResults1}. The results are presented in Fig. \ref{Fig9} and Table \ref{TableResults2}. As in Table \ref{TableResults1}, TRF+HFP+LMS provides the biggest NAUC, whereas HFP+LMS provides the smallest EFP.
	
	\begin{table}[h]
		\setlength{\tabcolsep}{0.5em} 
		\renewcommand{\arraystretch}{1.1}
		\centering
		\caption{NAUC and EFP metrics for the VC method trained on VC-PRG-1:5 and tested on VC-PRG-6.}
		\begin{tabular}{lcc}
			\hline\hline
			\multicolumn{1}{c}{Features} & 
			\multicolumn{1}{c}{NAUC} & \multicolumn{1}{c}{EFP [\%] ($ \Delta_{\text{EFP}}$ [\%])} \\
			\hline\hline
			All features & $0.702$ & $8.45$ ($0.34$) \\
			TRF+HFP+LMS & $\mathbf{0.729}$ & $6.72$ ($0.17$) \\
			HFP+LMS & $0.695$ & $\mathbf{6.55}$ ($0.17$) \\
			\hline\hline
		\end{tabular}
		\label{TableResults2}
	\end{table}

	With respect to the first experiment, detection thresholds corresponding to EFP changed from $ 78\% T_d $ to $ 74\% T_d $ (all features), $ 78\% T_d $ to $ 75\% T_d $ (TRF+HFP+LMS) and $ 81\% T_d $ to $ 79\% T_d $ (HFP+LMS). In that sense, HFP+LMS is characterized by the most stable behaviour.
	
\begin{figure}[thpb]
		\centering
		\includegraphics{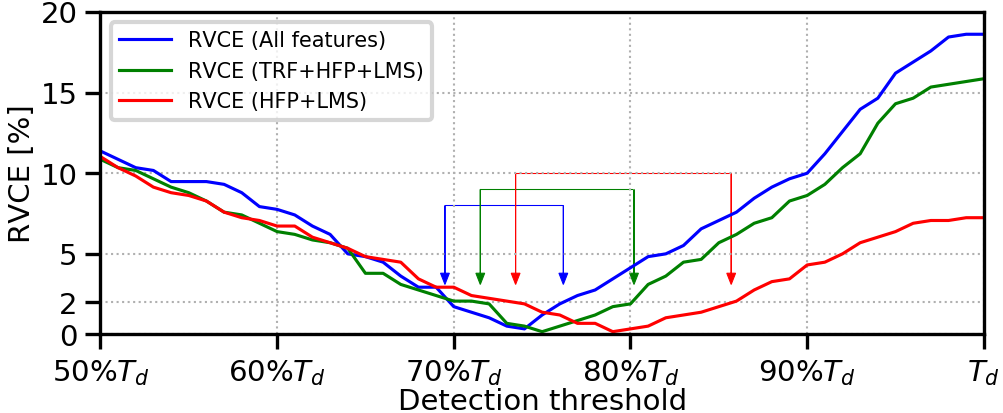}
		\caption{Relative VC error for the case when different datasets are used for training and testing. Arrows delimit intervals where RVCE is below $ 2 \% $.}
		\label{Fig10}
	\end{figure}

	In Fig. \ref{Fig10}, we present the RVCE values for detection thresholds within $ [50\% T_d,T_d] $. In terms of RVCE, HFP+LMS (red line) significantly outperforms the other two combinations. Namely, its RVCE is below $ 2 \% $ within $ [74\% T_d,85\% T_d] $, as opposed to $ [70\% T_d,76\% T_d] $ (all features, blue line) and $ [72\% T_d,80\% T_d] $ (TRF+HFP+LMS, green line). Arrows in Fig. \ref{Fig10} indicate the threshold ranges where $ \text{RVCE} < 2\% $. For the EFP thresholds obtained in the first experiment, $ \text{RVCE}=2.76\% $ (all features), $ 1.21\% $ (TRF+HFP+LMS) and $ 0.52\% $ (HFP+LMS). Difference in performance is much more striking with very high thresholds. For example, for the highest value of threshold, $ \text{RVCE}=18.62\% $ (all features), $ 15.86\% $ (TRF+HFP+LMS) and $ 7.24\% $ (HFP+LMS).
	
\begin{figure}[thpb]
		\centering
		\includegraphics{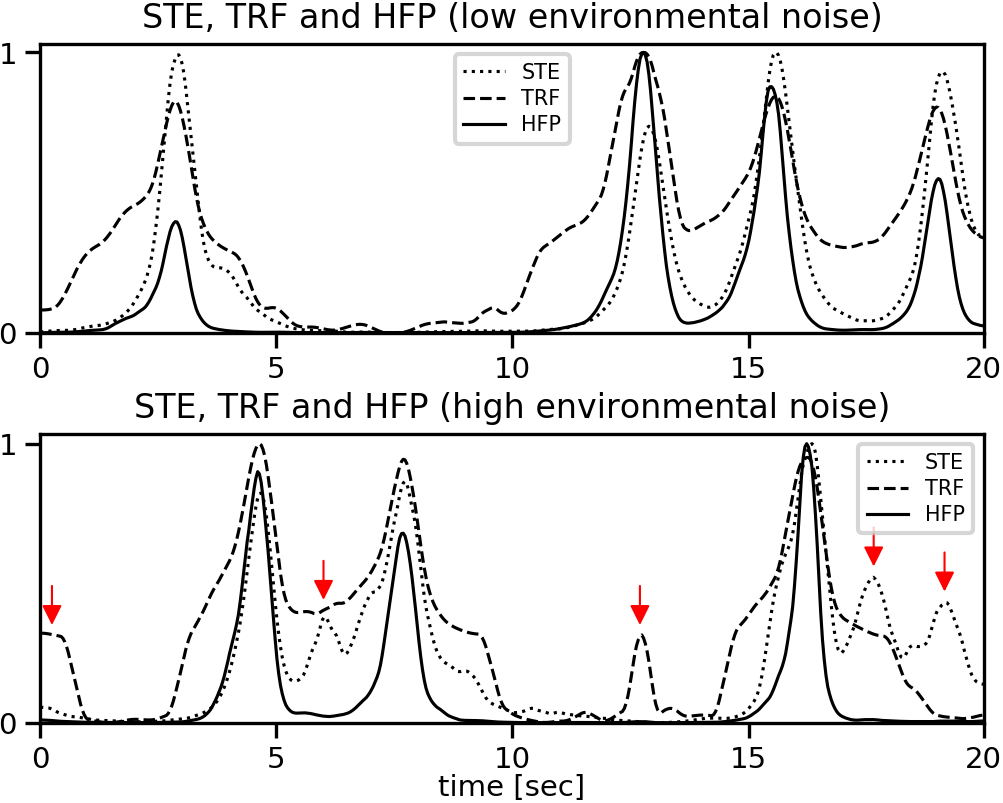}
		\caption{STE, TRF and HFP features in low-noise (top plot) and high-noise environment (bottom plot). Red arrows in the bottom plot indicate the positions of acoustic-noise (false) peaks.}.
		\label{Fig11}
	\end{figure}

	This experiment stresses the importance of introducing the HFP feature in the case of higher environmental noise. By analyzing the behaviour of the 1-D features (STE, TRF and HFP), we may conclude that HFP is much more robust to environmental noise than STE and TRF, i.e., it produces less false peaks than the latter two. This is illustrated in Fig. \ref{Fig11}, which depicts these features of two sound files recorded at the same location, one with low (top plot) and the other with high environmental noise (bottom plot). In addition to being robust to environmental noise, the HFP peaks are narrower than those of STE and TRF, which provides a capability of improved separation of close vehicles.
	
	\section{CONCLUSIONS}\label{Conclusions}
	
	We proposed a method for sound-based vehicle counting in low-to-moderate traffic flow. The method is based on a novel vehicle-to-microphone distance function which allows us to distinguish between the sounds due to vehicles passing by the microphone and all other sounds. The proposed distance is predicted using standard acoustic features and one novel feature which improves prediction robustness in noisy environments. The robustness is manifested in a very low counting error within a wide range of detection threshold values when testing is performed for a traffic location not considered in training.
	
	The future research will aim to identify and resolve error-causing situations in the proposed method, such as sound occlusion due to vehicles passing by the microphone simultaneously. We will also consider other regression approaches, such as recurrent neural networks, which enable accurate modelling of the sequential data. Finally, other traffic monitoring issues, such as vehicle speed estimation, will be addressed.
	
	\balance
		
	\bibliographystyle{IEEEtran}
	\bibliography{References}
	
\end{document}